\documentclass[%
 reprint,
 superscriptaddress,
 showkeys,
nofootinbib,
 amsmath,amssymb,
 aps,
]{revtex4-2}

\usepackage{float}
\usepackage[caption=false]{subfig} 

\usepackage{graphicx}
\usepackage{dcolumn}
\usepackage{bm}
\usepackage[hypertexnames=true,colorlinks=true,urlcolor=blue,bookmarks=true,citecolor=blue,breaklinks=true,pdftex]{hyperref}

\begin{document}

\preprint{APS/123-QED}

\title{Partially Constrained Internal Linear Combination: a method for low-noise CMB foreground mitigation}

\author{Y.~Sultan~Abylkairov}%
 \email{sultan.abylkairov@nu.edu.kz}
\affiliation{Physics Department, Nazarbayev University, Nur-Sultan, Kazakhstan}
\affiliation{Energetic Cosmos Laboratory, Nazarbayev University, Nur-Sultan, Kazakhstan}
\affiliation{Department of Applied Mathematics and Theoretical Physics, University of Cambridge, CB3 0WA, UK}
\author{Omar~Darwish}
\affiliation{Department of Applied Mathematics and Theoretical Physics, University of Cambridge, CB3 0WA, UK}
\author{J.~Colin~Hill}
\affiliation{Department of Physics, Columbia University, New York, NY, USA 10027}
\affiliation{Center for Computational Astrophysics, Flatiron Institute, New York, NY, USA 10010}
\author{Blake~D.~Sherwin}
\affiliation{Department of Applied Mathematics and Theoretical Physics, University of Cambridge, CB3 0WA, UK}
\affiliation{Kavli Institute for Cosmology, University of Cambridge, Madingley Road, Cambridge CB3 OHA, UK}

\date{\today}

\begin{abstract}
Internal Linear Combination (ILC) methods are some of the most widely used multi-frequency cleaning techniques employed in CMB data analysis. These methods reduce foregrounds by minimizing the total variance in the coadded map (subject to a signal-preservation constraint), although often significant foreground residuals or biases remain.  A modification to the ILC method is the constrained ILC (cILC), which explicitly nulls certain foreground components; however, this foreground nulling often comes at a high price for ground-based CMB datasets, with the map noise increasing significantly on small scales. In this paper we explore a new method, the partially constrained ILC (pcILC), which allows us to optimize the tradeoff between foreground bias and variance in ILC methods. In particular, this method allows us to minimize the variance subject to an inequality constraint requiring that the constrained foregrounds are reduced by at least a fixed factor, which can be chosen based on the foreground sensitivity of the intended application. We test our method on simulated sky maps for a Simons Observatory-like experiment; we find that for cleaning thermal Sunyaev-Zel'dovich (tSZ) contamination at $\ell \in [3000,4800]$, if a small tSZ residual of 20\% of the standard ILC residual can be tolerated, the variance of the CMB temperature map is reduced by at least 50\% over the cILC value. We also demonstrate an application of this method to reduce noise in CMB lensing reconstruction.
\end{abstract}

\maketitle


\section{\label{sec:intro}Introduction}

The Cosmic Microwave Background radiation (CMB) is one of our most important sources of information about cosmology and fundamental physics. Over the past decades, much of its constraining power has arisen from the primary CMB anisotropies. However, increasingly, the CMB is also being used as a backlight to understand the distribution of matter, gas and tracers lying between us and the last scattering surface, using the secondary anisotropies these imprint into the microwave background.

To robustly analyze either the primary CMB or the individual astrophysical contributions, multifrequency \emph{component separation methods}, which use different frequency dependences to disentangle the different components, are becoming increasingly important.

There are several methods that have been proposed to separate the CMB signal, or another astrophysical signal of interest, from the other components that are present in an observed CMB map. Perhaps the most widely used method is the Internal Linear Combination (ILC; \cite{Bennett_2003,Tegmark_2003,Eriksen_2004,2009A&A...493..835D,2014A&A...571A..21P,2014A&A...571A..12P,2016A&A...594A..22P,2020PhRvD.102b3534M,2020A&A...641A...4P}) method, which combines in a linear fashion multi-frequency observations in order to extract an unbiased estimate of the desired component (e.g. CMB)\footnote{Note because of frequency dependence, these frequency methods in general will extract a combination of CMB+kSZ+other frequency independent elements.}. This method employs a linear combination of frequency channels that minimizes the total map variance, subject to the constraint of an unbiased recovery of the desired component, with weights calculated from an empirically determined covariance matrix. A frequently-used extension of ILC is the constrained ILC (cILC; \cite{Remazeilles_2011}), where the linear combination is constructed in such a way to minimize the variance subject to the additional constraint that a particular component, with known spectral dependence, is nulled in the extracted map.  Without such nulling (also known as ``deprojection''), ILC maps can possess significant foreground residuals~(e.g.,~\cite{Madhavacheril_2018,2018MNRAS.479.4239C}).  The resulting deprojected maps have a wide range of applications, including primordial non-Gaussianity~\cite{Hill_2018,Planck2018NG}, cross-correlations (e.g., studying kSZ~\cite{Hill_2016,Planck2018kSZ} or ISW~\cite{Planck2018isotropy}), CMB lensing reconstruction~\cite{Madhavacheril_2018,darwish2020atacama}, and primordial B-modes~\cite{Remazeilles_2020}.  However, a major downside of using constrained ILC methods is that the additional constraints often lead to a substantial noise increase in the resulting map, particularly for ground-based CMB experiments with a moderate number of frequency channels.

However, this large noise penalty is not generally necessary if we only wish to obtain a map with a reduced level of foregrounds. Depending on the application, the complete nulling of foreground contamination may not be required, and it may instead be sufficient to merely reduce the contamination by a large factor in amplitude. This is the goal of this paper, which presents the partially constrained ILC (pcILC) method. This method guarantees an overall foreground bias that is reduced by at least a fixed factor, but, by not requiring foregrounds to be completely nulled, can result in a significantly lower noise than the cILC. The method is easily applicable; as an example, we will show an application to foreground cleaning for CMB gravitational lensing reconstruction.

In Section~\ref{sec:method}, following a brief review of existing multi-frequency component separation methods, we will introduce the pcILC. In Section~\ref{sec:results} we will show and discuss the results of our method when applied to simulations, and to an example of CMB lensing reconstruction. We conclude in Section~\ref{sec:concl}.  Further technical results are collected in the appendices.

\section{METHOD}
\label{sec:method}

Obtaining a clean and accurate CMB map from observational data is difficult due to various foreground signals such as the thermal Sunyaev-Zel’dovich (tSZ) effect and the Cosmic Infrared Background (CIB). In this section, we will quickly review the standard ILC, the constrained ILC, and finally our new proposed method, the partially constrained ILC. 

\subsection{The ILC method}
The ILC is a commonly used method because it requires minimal modeling assumptions about the data and has considerable flexibility in the choice of domain in which to extract the signal of interest. If we have $N_{\nu}$ frequency channels in our observational data, then for each pixel $p$ we can write an $N_{\nu}\times 1$ vector where each row represents the observed map at the corresponding frequency channel:
\begin{equation}
    \mathbf{y}(p)=\mathbf{a}s(p)+\mathbf{A}_f\mathbf{s}_f(p)+\mathbf{n}(p)
    \label{ILC_map_model}
\end{equation}
where $\mathbf{a}$ is the spectral energy distribution (SED) response vector of the desired signal $s(p)$, $\mathbf{A}_f$ is the mixing matrix for the foreground components $\mathbf{s}_f$ (to know how much a specific foreground $i$ contributes to the map at the observed frequency $j$), and $\mathbf{n}(p)$ is the noise.  We will focus solely on the case of CMB ILC reconstruction, and thus $\mathbf{a}$ is the CMB SED, which is unity when working in thermodynamic temperature units.  Note that $p$ can be a point in any desired space, e.g., in harmonic space, real space, or a needlet frame.  The ILC solution provides a linear combination of maps $\hat{s}= \mathbf{w}^T\mathbf{y}$ that recovers the component of interest, in this case the CMB, and has a minimum variance
\begin{equation}
    \mathrm{min}\left(\langle \hat{s}^2 \rangle-\langle\hat{s} \rangle^2\right)=\mathrm{min}(\mathbf{w}^T\mathbf{R}\mathbf{w})
    \label{eq:StandILC,var_zero}
\end{equation}
where $\mathbf{R}=\langle \mathbf{y}\mathbf{y}^T \rangle-\langle \mathbf{y} \rangle \langle \mathbf{y}^T \rangle$ is the covariance matrix of the data. Solving Eq. (\ref{eq:StandILC,var_zero}) under the constraint $\mathbf{w}^T\mathbf{a} = 1$, to ensure an unbiased recovery of the component of interest, gives the ILC weights (e.g.,~\cite{Eriksen_2004}):
\begin{equation}
    \mathbf{w}^T_{\text{ILC}}=\big(\mathbf{a}^T\mathbf{R}^{-1}\mathbf{a}\big)^{-1}\mathbf{a}^T\mathbf{R}^{-1}
\label{solution:Standard}
\end{equation}

\subsection{The constrained ILC method}
The cILC similarly involves building a linear combination of observed maps, at different frequencies, $\hat{s}= \mathbf{w}^T\mathbf{y}$ that recovers the component of interest with minimum possible variance; however, the cILC involves the additional constraint of nulling some unwanted foreground or other components.

To recover the CMB signal, while deprojecting some foregrounds, i.e., nulling some components of the $\mathbf{s}_f$, we use the corresponding SED vectors from the mixing matrix $\mathbf{A}_f$. We define these SED vectors as ($\mathbf{b_1},\mathbf{b_2},...,\mathbf{b_m}$). Then we can write a condition under which we completely deproject these components:
\begin{equation}
\begin{cases} 
\mathbf{w}^T\mathbf{b_1}=0\\
\mathbf{w}^T\mathbf{b_2}=0\\
\vdots\\
\mathbf{w}^T\mathbf{b_m}=0
\end{cases}
\end{equation}
In this way we guarantee that the contribution of the selected foregrounds to the final linear combination map will be zero, which is not necessarily true for the ILC case.  However, this deprojection comes at a price: since we have used one or more degrees of freedom for the deprojection, the noise in the final cILC map is guaranteed to be higher than that in the standard ILC case.

Formally, the constrained ILC solution provides a linear combination of maps $\hat{s}= \mathbf{w}^T\mathbf{y}$ such that it has minimal variance, subject to the constraints $\mathbf{w}^T\mathbf{a} = 1$, $\mathbf{w}^T\mathbf{b_1}=0$, $\mathbf{w}^T\mathbf{b_2}=0$,...,$\mathbf{w}^T\mathbf{b_m}=0$. In this case, the weights are (e.g.,~\cite{Remazeilles_2011}):
\begin{equation}
    \mathbf{w}^T_{\text{cILC}}=\mathbf{e}^T\big(\mathbf{A}^T\mathbf{R}^{-1}\mathbf{A}\big)^{-1}\mathbf{A}^T\mathbf{R}^{-1}
\end{equation}
where $\mathbf{A}=[\mathbf{a}$ $\mathbf{b_1}\cdots\mathbf{b_m}]$ is a matrix of size $N_{\nu} \times (m+1)$, and $\mathbf{e}^T=[1$ $0\cdots0]$ is a vector of $1\times(m+1)$, so that we can recover the CMB (which, in our formalism, is always the first component) if the unwanted components are foregrounds.


\subsection{The partially constrained ILC method}
\label{subsec:pcILC_method}

In the constrained ILC, by deprojection we lose one degree of freedom for each deprojected component; this inevitably leads to an increase of variance in the combined map. Here we propose a new method where we partially deproject foregrounds to get an intermediate solution between the ILC and the constrained ILC, i.e., to achieve a balance between foreground bias and variance reduction. 

As a starting point, suppose we have just the CMB and one foreground component, with the SED vector $\mathbf{b_1}$, that we wish to reduce in the final combination. Partial deprojection can be expressed as
\begin{equation}
|\mathbf{w}^T\mathbf{b_1}|\leq \epsilon\ ,
\label{eq:epsilon}
\end{equation}
where $\epsilon$ is some arbitrary positive number which controls the level of residual foregrounds in the final map. By defining new ``slack variables'' $s_1$ and $s_2$ to turn inequality constraints to equality constraints, we write the modulus in Eq. (\ref{eq:epsilon}) as two equations with different signs~\cite{LagrangeIneq}; the inequality constraint can then be expressed as follows:

\begin{equation}
\begin{split}
    \epsilon-\mathbf{w}^T\mathbf{b_1}-s_1^2=0\\
    \epsilon+\mathbf{w}^T\mathbf{b_1}-s_2^2=0
\end{split}
\end{equation}
To find weights $\mathbf{w}$ such that the combined map has minimal variance under constraints, we use the method of Lagrange multipliers:
\begin{equation}
\begin{split}
    \mathcal{L}=\mathbf{w}^T\mathbf{R}\mathbf{w}+\lambda(1-\mathbf{w}^T\mathbf{a})+\lambda_1(\epsilon-\mathbf{w}^T\mathbf{b_1}-s_1^2)&\\
    +\lambda_2(\epsilon+\mathbf{w}^T\mathbf{b_1}-s_2^2)
\label{eq:Lagrange}
\end{split}
\end{equation}
Minimizing this, we obtain a linear system of equations:
\begin{align}
\begin{cases}
    \frac{\partial \mathcal{L}}{\partial \mathbf{w}^T}=2\mathbf{R}\mathbf{w}-\lambda \mathbf{a}-\lambda_1 \mathbf{b_1}+\lambda_2 \mathbf{b_1}=\mathbf{0}\\
    \frac{\partial \mathcal{L}}{\partial s_1}=-2\lambda_1 s_1=0\\
    \frac{\partial \mathcal{L}}{\partial s_2}=-2\lambda_2 s_2=0\\
    \frac{\partial \mathcal{L}}{\partial \lambda}=1-\mathbf{a}^T\mathbf{w}=0\\
    \frac{\partial \mathcal{L}}{\partial \lambda_1}=\epsilon-\mathbf{b_1}^T\mathbf{w}-s_1^2=0\\
    \frac{\partial \mathcal{L}}{\partial \lambda_2}=\epsilon+\mathbf{b_1}^T\mathbf{w}-s_2^2=0\\
\end{cases}
\label{eq:syst_of_eq}
\end{align}
We will now outline a step-by-step solution of the system of Eq. (\ref{eq:syst_of_eq}):\\
\textit{Step one}: The first equation of the system gives
\begin{equation}
    \mathbf{w}=\frac{1}{2}\mathbf{R}^{-1}(\lambda \mathbf{a}+\lambda_1 \mathbf{b_1}-\lambda_2 \mathbf{b_1}).
    \label{eq:from_sys_w}
\end{equation}
\textit{Step two}: substituting Eq. (\ref{eq:from_sys_w}) into other equations that contain $\mathbf{w}$ in the system of equations
\begin{equation}
    \mathbf{a}^T \frac{1}{2}\mathbf{R}^{-1}(\lambda \mathbf{a}+\lambda_1 \mathbf{b_1}-\lambda_2 \mathbf{b_1}) = 1
    \label{eq:from_sys_1}
\end{equation}
\begin{equation}
    \mathbf{b_1}^T \frac{1}{2}\mathbf{R}^{-1}(\lambda \mathbf{a}+\lambda_1 \mathbf{b_1}-\lambda_2 \mathbf{b_1}) = \epsilon - s_1^2
    \label{eq:from_sys_2}
\end{equation}
\begin{equation}
    \mathbf{b_1}^T \frac{1}{2}\mathbf{R}^{-1}(\lambda \mathbf{a}+\lambda_1 \mathbf{b_1}-\lambda_2 \mathbf{b_1}) = s_2^2 - \epsilon 
    \label{eq:from_sys_3}
\end{equation}
\textit{Step three}: from the second and third equations of the system of equations it follows: $\lambda_1=0$ or $s_1=0$ or $\lambda_1 = 0\  \&\  s_1=0$, and $\lambda_2=0$ or $s_2=0$ or $\lambda_2 = 0\ \&\  s_2=0$. By simply substituting the possible combinations into Eq. (\ref{eq:from_sys_1}-\ref{eq:from_sys_3}), we can build the following table:

\begin{table}[H]
\caption{Table showing which variables are zero, out of all the combinations described in step three, and whether for this combination of zero-valued variables it is possible (Yes or No) to find a solution for the system of equations in Eq. (\ref{eq:syst_of_eq}).}
\label{table:pcILC_solutions}
\begin{tabular}{ |p{0.7cm}|p{0.7cm}|p{1cm}|p{0.7cm}|p{0.7cm}|p{1cm}|p{0.7cm}|p{0.7cm}|p{1cm}| }
 \hline
 \multicolumn{3}{|c}{$\lambda_1$} &
 \multicolumn{3}{|c|}{$s_1$} &
 \multicolumn{3}{c|}{$\lambda_1\& s_1$} \\
 \hline
 $\lambda_2$  & $s_2$ & $\lambda_2\& s_2$ & $\lambda_2$ & $s_2$  & $\lambda_2\& s_2$ & $\lambda_2$ & $s_2$ & $\lambda_2\& s_2$\\
 \hline
 Yes  & Yes & No & Yes & No & No & No & No & No\\
 \hline
\end{tabular}
\end{table}

Table \ref{table:pcILC_solutions} shows all possible combinations of different variables being zero: $\lambda_1=0$ or $s_1=0$ or both, and $\lambda_2=0$ or $s_2=0$ or both. Specifically, the first row labels which of the $\lambda_1, s_1$ variables are zero (or whether both are zero); the second row similarly labels which of the $\lambda_2, s_2$, $\lambda_2\& s_2$ variables are zero, and the third row shows whether this combination of zero-valued variables allows the system of equations to have an answer (Yes/No). \\
\textit{Step four}: Finally, using simple algebra for $\lambda_1=0$ and $\lambda_2=0$ we have the standard ILC solution, for $\lambda_1=0$ and $s_2=0$ we have $\mathbf{w}_{\text{pcILC}-}$ (see Eq. (\ref{solution:partial_const})), and for $\lambda_2=0$ and $s_1=0$ we have $\mathbf{w}_{\text{pcILC}+}$ (see Eq. (\ref{solution:partial_const})).

Therefore, solving the linear system of Eq. (\ref{eq:syst_of_eq}) gives us three solutions:
\begin{itemize}
    \item The first solution is the ILC solution.
    
    \item The other two take values at the boundary, i.e., $\mathbf{w}^T_{\text{pcILC}_{\pm}}\mathbf{b_1}=\pm \epsilon$, and have the following form:
\end{itemize}
\begin{equation}
    \mathbf{w}^T_{\text{pcILC}_{\pm}}=\frac{\mathbf{a}^T\mathbf{R}^{-1}(K_{b} \mp K_{ab}\epsilon)+\mathbf{b_1}^T\mathbf{R}^{-1}(\pm K_{a}\epsilon-K_{ab})}{K_{a}K_{b} - K_{ab}^2}
    \label{solution:partial_const}
\end{equation}
where $K_{a}=\mathbf{a}^T\mathbf{R}^{-1}\mathbf{a}$, $K_{b}=\mathbf{b_1}^T\mathbf{R}^{-1}\mathbf{b_1}$ and $K_{ab}=\mathbf{a}^T\mathbf{R}^{-1}\mathbf{b_1}$.

The answers above make sense if we consider this problem as follows: The variance function $\mathbf{w}^T\mathbf{R}\mathbf{w}$: $\mathbb{R}^n\to \mathbb{R}$, where $n=N_{\nu}$ is a number of frequency channels.  Geometrically this is an elliptic paraboloid with a minimum at $\mathbf{w}=\mathbf{0}$. With our first constraint $\mathbf{w}^T\mathbf{a}=1$, our working domain is reduced to $\mathbb{R}^{n-1}$ and geometrically it is still an elliptic paraboloid with minimum at $\mathbf{w}_{\text{ILC}}$. Adding the inequality constraint $|\mathbf{w}^T\mathbf{b_1}|\leq \epsilon$ is equivalent to considering an allowed interval $I$ in $\mathbb{R}^{n-1}$. Therefore, the minimum will be the standard ILC solution if the interval includes $\mathbf{w}_{\text{ILC}}$, otherwise it will lie at the boundaries (i.e., if $\mathbf{w}_{\text{ILC}}\notin I$).

To find the correct overall solution, we first have to check if the inequality condition is satisfied for the standard ILC solution, i.e., $|\mathbf{w}^T_{\text{ILC}}\mathbf{b_1}|\leq \epsilon$. If so, then the answer ($\mathbf{w}_{\text{pcILC}}$) is equal to the standard ILC solution. Otherwise, we calculate the variance of the combined map for both weights $\mathbf{w}_{\text{pcILC}_{\pm}}$, i.e., $\mathbf{w}^T_{\text{pcILC}_{\pm}}\mathbf{R}\mathbf{w}_{\text{pcILC}_{\pm}}$, and compare them. The answer in this case is the weight vector for which we get the smallest variance. 

In the Appendix we discuss how to generalize this derivation to multiple partially deprojected components.

\begin{figure*}
\includegraphics[width=0.95\textwidth]{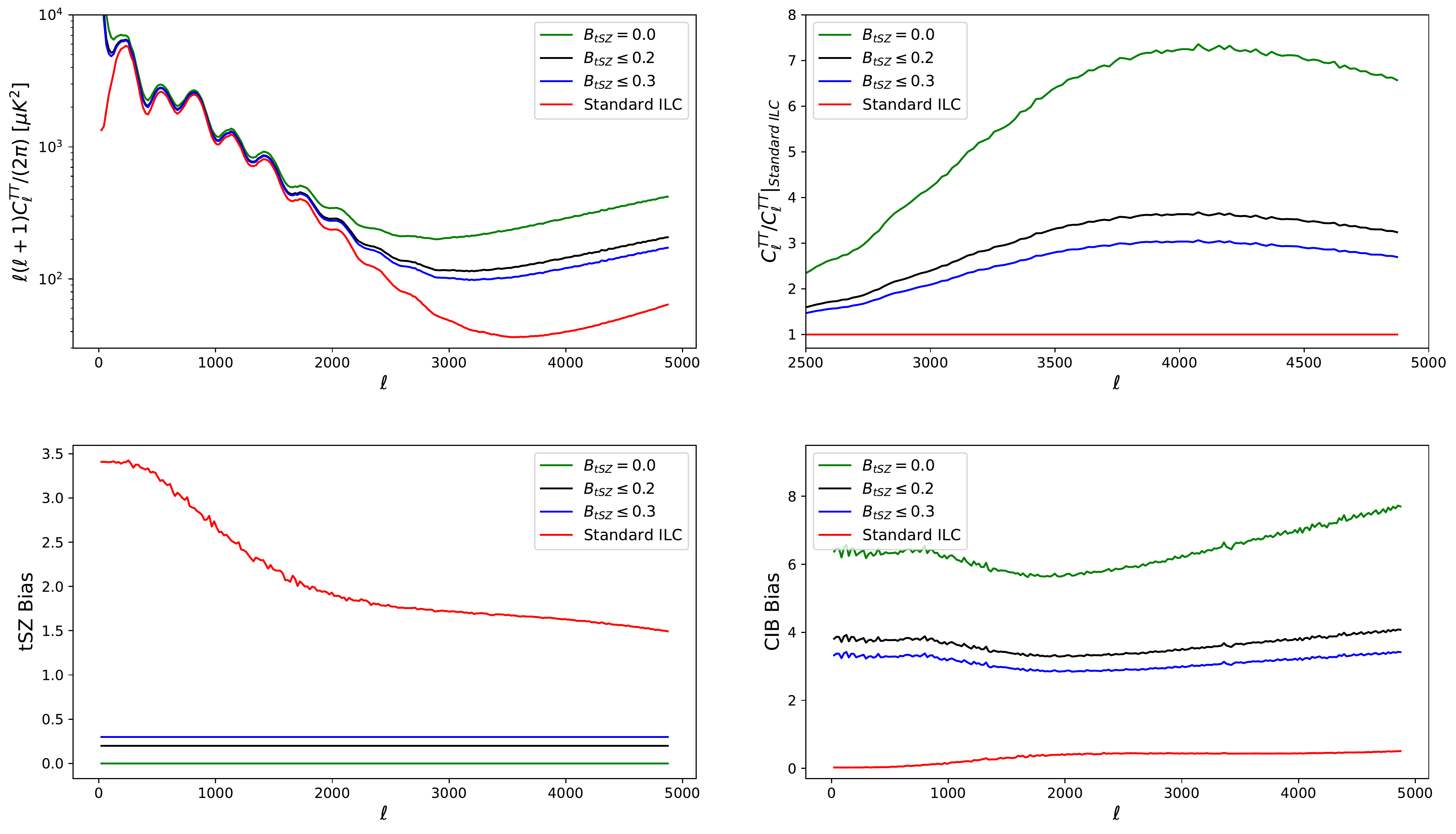}
\caption{\textbf{Partial deprojection of the tSZ component:} Results from our new pcILC method applied to simulated sky maps with various foreground bias threshold values as defined in Eq.~(\ref{bias_eq}) (blue and black curves) compared to the standard ILC (red) and the constrained ILC, or cILC, (green) results.  Upper left: total power spectra of the reconstructed (pc)ILC CMB maps.  Upper right: The ratio of the power spectra to the total CMB power spectrum obtained with the standard ILC.  Lower left and lower right: the residual tSZ power (left) and CIB power (right) in the coadded maps, measured relative to the power of the tSZ or CIB at 145 GHz (see Eq.~(\ref{bias_eq})).  It can be seen that, if a small bias can be tolerated, the pcILC method provides a significant variance reduction when compared with the cILC.}
\label{FIG:sz}
\end{figure*}

\begin{figure*}
\includegraphics[width=0.95\textwidth]{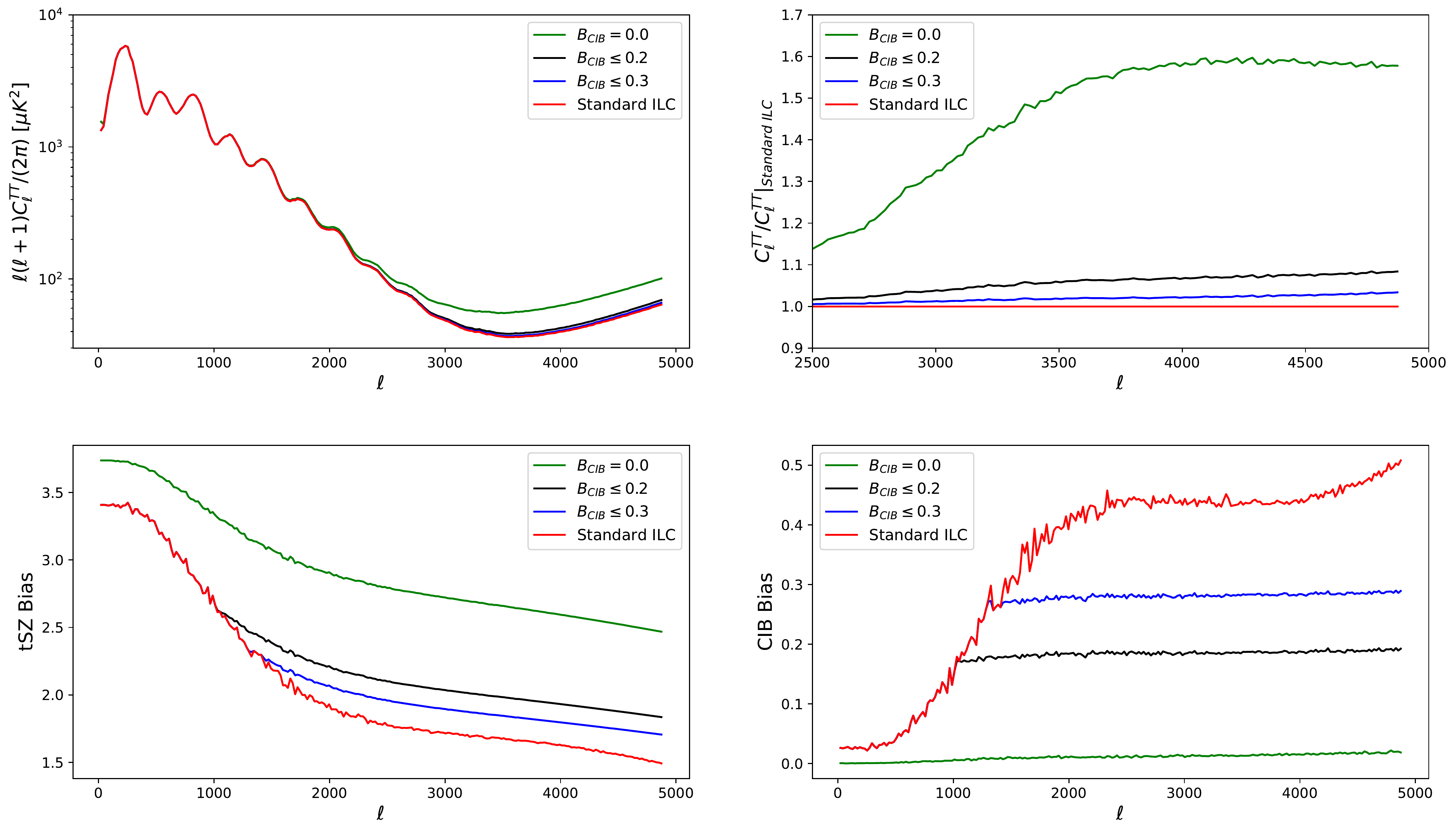}
\caption{\textbf{Partial deprojection of the CIB component:} The same configuration of plots as in Figure~\ref{FIG:sz}, but now with the CIB foreground component deprojected or partially deprojected, as labeled in the plot legends.  The CIB SED is taken to be a modified blackbody (see Eq.~(\ref{eq: SED_CIB})) in the ILC constraints, but the sky simulations are constructed with a realistic model that produces decorrelation and a non-rigid SED that varies with frequency and sky position.  This is why the residual CIB bias is slightly non-zero even when we set $B_{CIB} = 0$ (see the green curve in the lower right panel).}
\label{FIG:ci}
\end{figure*}

\begin{figure*}
\includegraphics[width=0.95\textwidth]{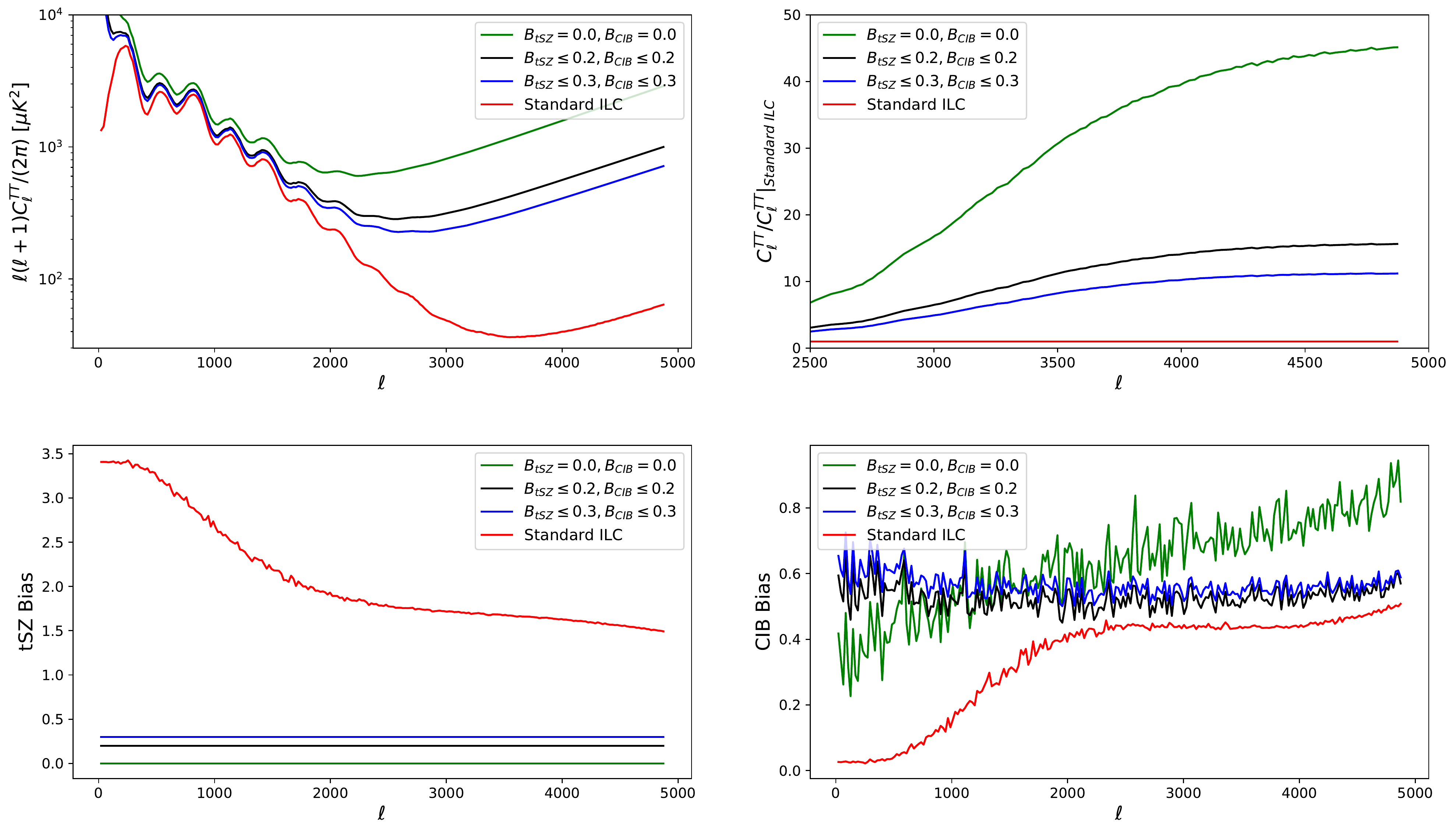}
\caption{\textbf{Partial deprojection of both tSZ and CIB components:} The same configuration of plots as in Figure~\ref{FIG:sz}, but with simultaneous deprojection or partial deprojection of both the tSZ and CIB foreground components, as labeled in the plot legends. Again, it can be seen in the top right panel that, if a small bias can be tolerated, the pcILC method provides a significant variance reduction when compared with the cILC. However, the bottom right panel shows that the CIB bias remains significantly non-zero even when we attempt to fully deproject this component.  This situation can arise when (partially) deprojecting multiple foregrounds with a small number of frequency channels, such that even small inaccuracies in the modeling lead to non-negligible residual foreground biases (see text for further discussion).}
\label{FIG:tw}
\end{figure*}

\section{\label{sec:results}RESULTS and Discussion}

\subsection{Simulations}
\label{subsec:sims}

We test our proposed method on the high-resolution simulations of the microwave sky\footnote{The simulations can be found at \url{https://lambda.gsfc.nasa.gov/simulation/tb_sim_ov.cfm}} generated by The Simons Observatory Collaboration \cite{SO_Ade_2019,Sehgal_2010}. For simplicity, we perform the ILC in harmonic space, but the novel aspects of our formalism can be straightforwardly applied in pixel space or on a needlet frame as well.  The simulation maps are constructed for six frequency channels at which the Simons Observatory (SO) will operate: 27 GHz, 39 GHz, 93 GHz, 145 GHz, 225 GHz, and 280 GHz.  For simplicity, delta-function passbands are assumed.

In this work, we use a simple sky model that includes the lensed CMB signal, the tSZ effect, and the CIB.  The tSZ effect is the inverse-Compton scattering of CMB photons off hot, free electrons, which generates a unique spectral distortion in the mm-wave bands~\cite{1969Ap&SS...4..301Z}.  The CIB is the cumulative thermal emission from dust grains heated by starlight in galaxies over cosmic history.  The lensing, tSZ, and CIB fields were constructed in these simulations by post-processing a large $N$-body simulation with prescriptions for each observable~\cite{Sehgal_2010}.  All components are thus realistically correlated.  Further details on each individual component can be found in Refs.~\cite{Sehgal_2010,SO_Ade_2019}, including adjustments that were made to more closely match recent measurements of these fields.  The noise model in the simulations is generated from the properties of the planned SO surveys, i.e., the ``baseline'' level for the SO Large Aperture Telescope (LAT) with observed sky fraction $f_{\rm sky} = 40\%$ (see \cite{SO_Ade_2019}). Note that the noise maps are correlated at 27 and 39 GHz, 93 and 145 GHz, and at 225 and 280 GHz due to the atmospheric correlations for frequency channels in the same optics tube (see \cite{SO_Ade_2019}). We combine the lensed CMB, tSZ, CIB, and noise components for each frequency channel.

\subsection{Frequency dependence of components}

To apply the cILC and pcILC methods, we need to know the frequency response models of the components that we wish to deproject or partially deproject. In this work, the components we will focus on are the tSZ and CIB.

For the tSZ effect \cite{1970Ap&SS...7....3S, 1969Ap&SS...4..301Z, 2006NCimB.121..487N} the frequency dependence in thermodynamic CMB temperature units is given by:
\begin{equation}
    f_{\text{tSZ}}(\nu)=x\frac{e^x+1}{e^x-1}-4
\end{equation}
where $x=h \nu/(k_B T_{\text{CMB}})$.  In contrast to the tSZ effect, the CIB is not a single field that is rigidly rescaled across frequency channels according to a fixed SED.  However, as an approximation, we adopt the following modified blackbody SED for the CIB~\cite{2020PhRvD.102b3534M}:
\begin{equation}
    f_{\text{CIB}}(\nu)\propto \frac{\nu^{3+\beta}}{e^{h\nu/(k_BT_{\text{CIB}})}-1} \Bigg(\frac{dB(\nu, T)}{dT}\Bigg|_{T=T_{\text{CMB}}} \Bigg)^{-1}
    \label{eq: SED_CIB}
\end{equation}
where $\beta=1.2$, $T_{\text{CMB}}=24$ K, and $B(\nu, T)$ is the Planck function, needed here to convert from specific intensity to thermodynamic CMB temperature units.  We emphasize that the CIB component in the simulated sky maps is not generated assuming this SED, but rather from detailed post-processing of a lightcone from an $N$-body simulation, using semi-analytic star formation prescriptions.  Thus, the simulated CIB maps do not follow a single, rigid SED, and they exhibit realistic decorrelation across frequency channels~\cite{SO_Ade_2019}.

Finally, for our frequency channels, the CMB SED is a constant and equal to unity, since we work in thermodynamic CMB temperature units.

\subsection{Choosing a value for $\epsilon$}

To understand how to choose a reasonable threshold value for partial deprojection (see Eq. (\ref{eq:epsilon})), we will first explain the calculation of the foreground bias values. We define a foreground bias fraction as: 
\begin{equation}
    B:=\frac{\mathbf{w}^T\mathbf{F}\mathbf{w}}{F_{145\times 145}}
    \label{bias_eq}
\end{equation}
where $\mathbf{F}$ is the empirically determined frequency-frequency covariance matrix for this foreground and $F_{145\times 145}$ is the power spectrum of this foreground at 145 GHz obtained from the simulations. This bias variable $B$ represents the size of the residual foreground power after applying the pcILC method (equal to ${\mathbf{w}^T\mathbf{F}\mathbf{w}}$), relative to the original foreground power at 145 GHz. We can then define a positive number $B^{th}$ such that $B\leq B^{th}$, which defines a threshold value of the foreground bias.  Note that from Eq.~(\ref{bias_eq}) we obtain the foreground bias for the standard ILC, by inserting the standard ILC weights in the numerator.  Based on this result, we know a reasonable upper bound for the threshold bias value $B^{th}$, since any threshold bias value above the standard ILC bias will just reproduce the standard ILC weights, as discussed in Section \ref{subsec:pcILC_method}. Depending on how much we want to reduce the variance in the final ILC map, we can thus choose any value between the standard ILC foreground bias and zero.  Finally, we can calculate the threshold value to be used for the pcILC weights determination using the following equation:
\begin{equation}
    \epsilon=\sqrt{B^{th} \cdot F_{145\times 145} }
\end{equation}

In this paper, for simplicity, we use the same threshold value for all $\ell$, even if for some $\ell$ it exceeds the standard ILC bias. Specifically, we constrain the tSZ bias to be below a threshold $B_{tSZ}\leq 0.2$ or $0.3$ and CIB bias to be below a threshold $B_{CIB}\leq 0.2$ or $0.3$.

\subsection{ILC: results and discussion}
Figure \ref{FIG:sz} shows the results obtained using ILC, cILC, and pcILC for CMB map reconstruction on the SO-like simulations, where the constraints are applied only to the tSZ foreground component.  As mentioned previously, for the pcILC method, we show results for both $B_{tSZ}\leq 0.2$ and $B_{tSZ}\leq 0.3$. All calculations are performed in linearly-spaced multipole bins of width $\Delta \ell=21$.  The upper left (ul) panel shows the total power spectrum of the lensed CMB signal reconstructed by various methods, and the upper right (ur) panel shows the ratio of these power spectra to the power spectrum of the lensed CMB obtained using the standard ILC method.  The lower left (ll) and right (lr) panels show the tSZ bias and the CIB bias, as defined in Eq.~(\ref{bias_eq}).  The same configuration is shown in Figures \ref{FIG:ci} and \ref{FIG:tw}, but in Figure \ref{FIG:ci} the pcILC method is implemented for the CIB foreground component with $B_{CIB}\leq 0.2$ and $B_{CIB}\leq 0.3$, and in Figure~\ref{FIG:tw} the pcILC method is implemented for both foreground components simultaneously, i.e., tSZ and CIB with $B_{tSZ}\leq 0.2$ and $B_{CIB}\leq 0.2$, and $B_{tSZ}\leq 0.3$ and $B_{CIB}\leq 0.3$.

The results in Figure~\ref{FIG:sz} show that when the tSZ component is fully deprojected with the cILC, the variance of the reconstructed CMB map increases by more than a factor of six compared to the standard ILC for some multipoles, and the residual CIB power becomes much higher than its value for the standard ILC (as expected, since the constrained weights have less freedom to adjust to minimize CIB contamination).  However, with partial deprojection such that $B_{tSZ}\leq 0.2, 0.3$ using the pcILC, the variance and the CIB bias only increase by a moderate amount over the standard ILC results.  In particular, for $B_{tSZ}\leq 0.3$ the resulting total power spectrum is less than half the power spectrum of the cILC map for $\ell\in [3000,4800]$.

For the CIB deprojection in Figure~\ref{FIG:ci}, the improvements when using the pcILC method are not as striking as for tSZ deprojection.  Nevertheless, if we can tolerate a small residual CIB bias, we still can lower the effective power spectrum by tens of percent and reduce the residual tSZ bias substantially by using the pcILC method instead of the cILC, as shown in Figure~\ref{FIG:ci}.

The main disadvantage of single-component cILC and pcILC is that we do not control the bias of other foregrounds, and therefore these biases can become larger, as shown previously. By deprojection and partial deprojection for two or more components, we control the level of bias of multiple foregrounds, or two for the case of our sky simulations here. In this case, where we deproject both tSZ and CIB, cILC increases the variance more than forty times compared to the ILC for high $\ell$, as shown in Figure \ref{FIG:tw}. Using the pcILC method, we can significantly reduce the variance.

Although pcILC performs well at reducing variance, the CIB bias becomes more sensitive to decorrelation and suboptimal SED modeling as additional constraints are added. The small decorrelation and suboptimal SED model of the CIB signal affect the accuracy of the CIB bias removal, as shown in Figure \ref{FIG:tw} (lr). However, we would expect the same variance reduction performance for more accurate models (see next subsection); experiments with more frequency channels should, in addition, suffer less from bias residuals when (partially) deprojecting multiple components (see Ref.~\cite{Remazeilles_2020}). In Appendix A, we analyze in detail the problem of accurately reducing the CIB bias.  We also note that including additional CIB components via a moment expansion~\cite{2017MNRAS.472.1195C} could reduce the bias seen in Figure \ref{FIG:tw}, although this may be challenging for experiments with a relatively small number of frequency channels.

Aside from the CIB bias calculation, we have verified that the simulated results match our forecast performance well. This gives further confidence in our method and our pipeline performance.

\begin{figure}
\centering
\includegraphics[width=0.95 \linewidth]{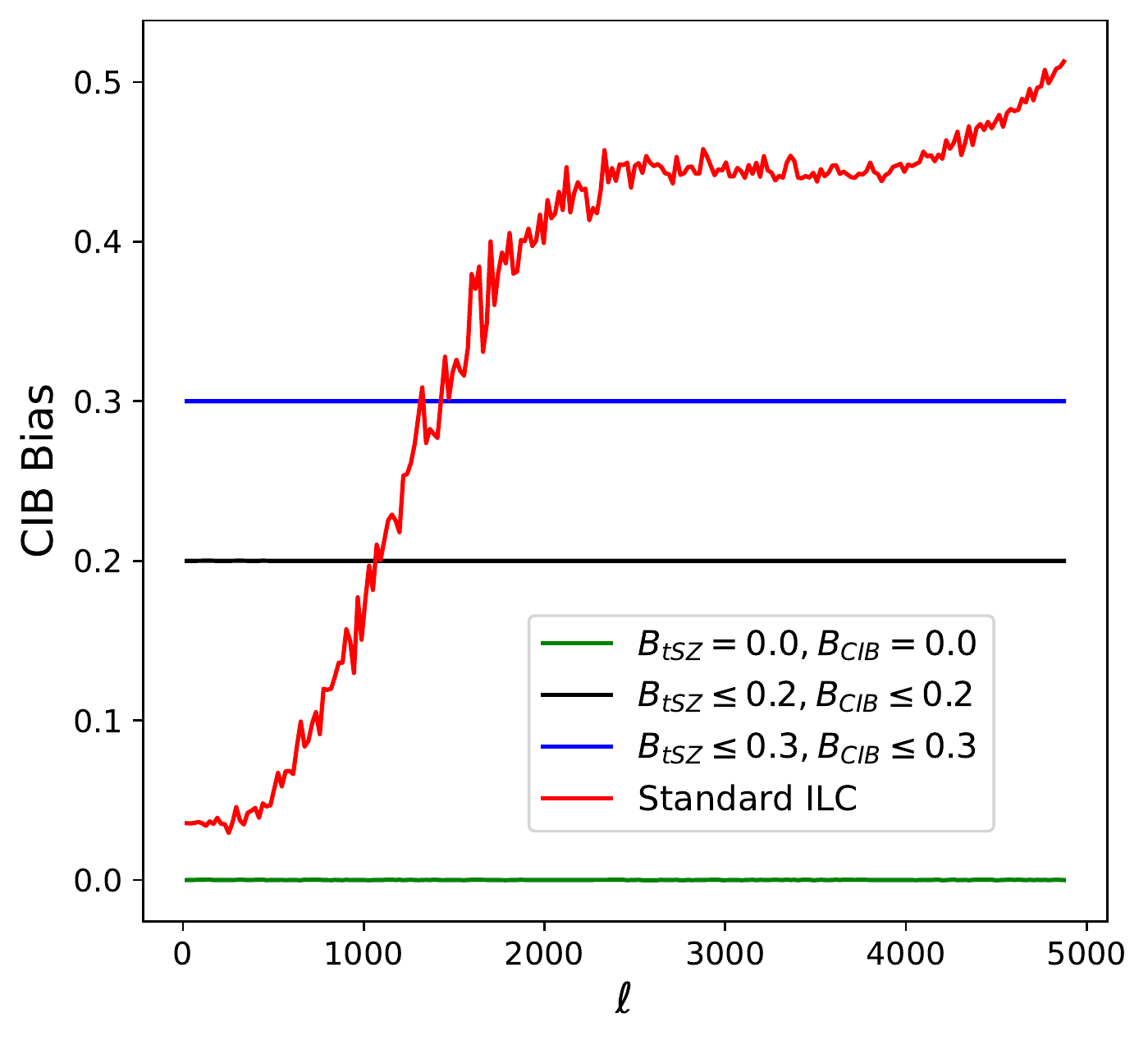}
\caption{\textbf{Partial deprojection of both tSZ and CIB components:} The residual CIB power in the coadded maps, measured relative to the power of the CIB at 145 GHz.  The results shown here are obtained for the simplified sky simulations described in Section~\ref{subsec:sims_simple}, in which the CIB field is comprised of a single component that is simply rescaled in frequency using Eq.~(\ref{eq: SED_CIB}).  In contrast to Figure~\ref{FIG:tw}, the CIB bias now behaves as expected, which demonstrates that the behavior seen previously was due to CIB decorrelation and SED variations.}
\label{FIG:tw_v2}
\end{figure}

\subsection{Validation of CIB results using simplified sky simulations}
\label{subsec:sims_simple}

In this subsection, we use a simplified version of the simulated sky described in Section~\ref{subsec:sims}, with the only difference that instead of the standard CIB signals, we use one CIB signal at 145 GHz and scale it to the other frequency channels using Eq.~(\ref{eq: SED_CIB}).  Thus, in these simplified simulations, the CIB field is comprised of a single component, which is rescaled across frequencies with a rigid SED.  We combine this component with the CMB, tSZ, and noise as done for the original simulations.  With these simplified sky maps, we check how well our method works for partially deprojecting both tSZ and CIB when we have a perfectly-understood, one-component CIB signal.  The results are shown in Figure~\ref{FIG:tw_v2}, which demonstrate that the CIB bias is now successfully removed when $B_{tSZ} = 0 = B_{CIB}$, and behaves as expected in the other cases shown.  This validates our claim that the residual CIB biases seen in Figure~\ref{FIG:tw} are indeed due to SED variations and decorrelation in the CIB signal in the original simulations.  Also note that for two-component partial deprojection, in order to obtain the standard ILC solution, the corresponding intervals from the inequality constraints for tSZ and CIB, that is $I_{\text{tSZ}}$ and $I_{\text{CIB}}$, must include $\mathbf{w}_{\text{ILC}}$ (i.e., $\mathbf{w}_{\text{ILC}} \in I_{\text{tSZ}}$ and $\mathbf{w}_{\text{ILC}} \in I_{\text{CIB}}$). This is why in Figure \ref{FIG:tw_v2}, the CIB bias may be higher than its corresponding bias from the standard ILC if at the same time the tSZ bias is lower than its corresponding bias from the standard ILC.

\begin{figure}
    \centering
    \includegraphics[width=0.95 \linewidth]{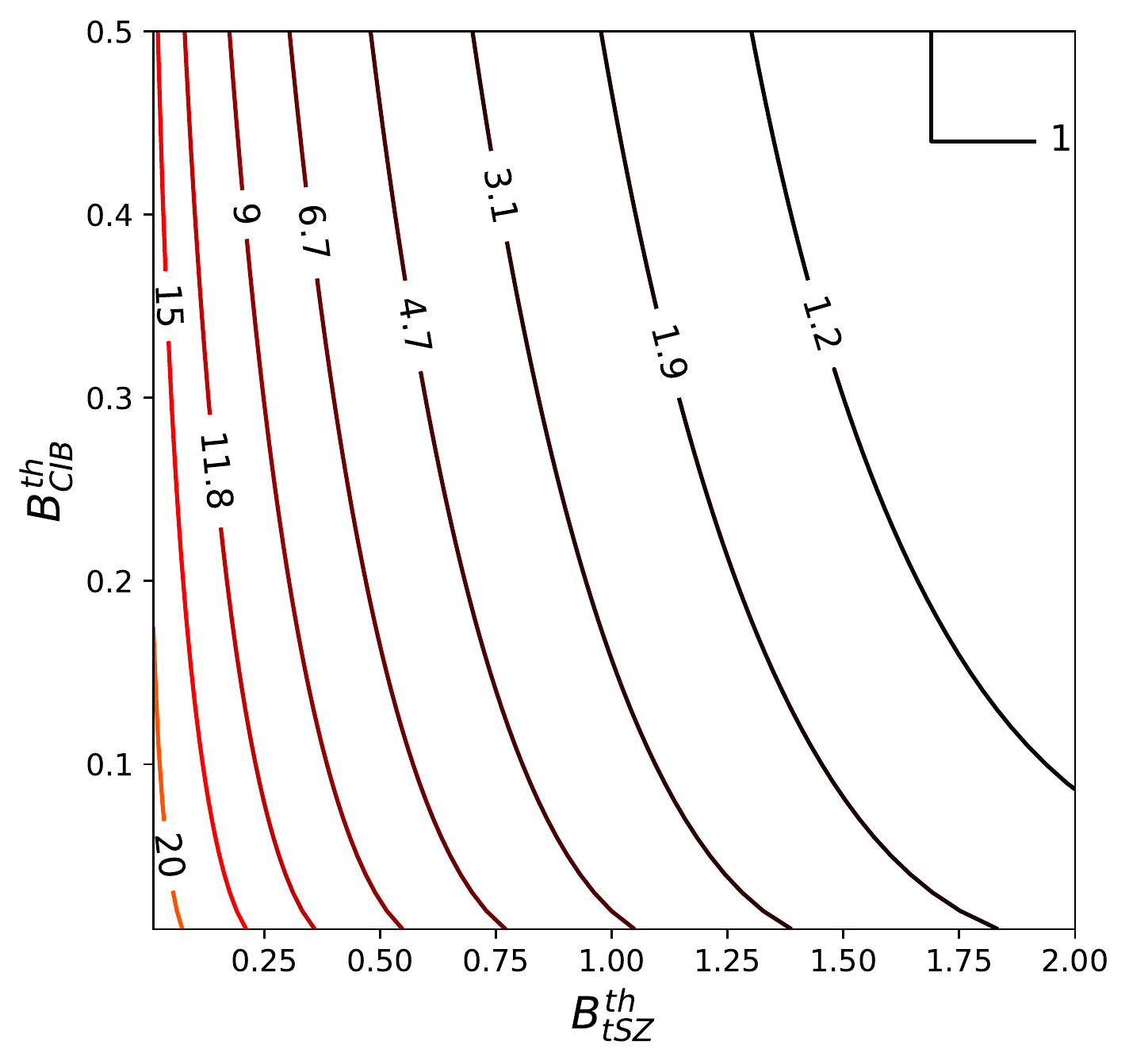}
    \caption{Contour plot of the ratio of the total CMB map power spectra obtained with pcILC for two components (tSZ and CIB) to the spectrum of the total CMB obtained with standard ILC, at $\ell=3500$ as a function of $B^{th}_{tSZ}$ and $B^{th}_{CIB}$.}
    \label{fig:conture_plot}
\end{figure}

Figure~\ref{fig:conture_plot} shows how the ratio of the power spectra obtained with pcILC for two components (tSZ and CIB) to the spectrum of the total lensed CMB obtained with standard ILC at $\ell = 3500$ varies with $B^{th}_{tSZ}$ and $B^{th}_{CIB}$.  Note that when $B^{th}_{tSZ}$ and $B^{th}_{CIB}$ are equal to or greater than the corresponding standard ILC bias values, the ratio becomes equal to one, as expected, since the standard ILC weights are recovered in this case.

\begin{figure*}
\centering
\subfloat[Subfigure 1 caption:][Case with $\ell_{\rm{max}}=3000$ for reconstruction with all estimators.]{
\includegraphics[width=0.46\linewidth]{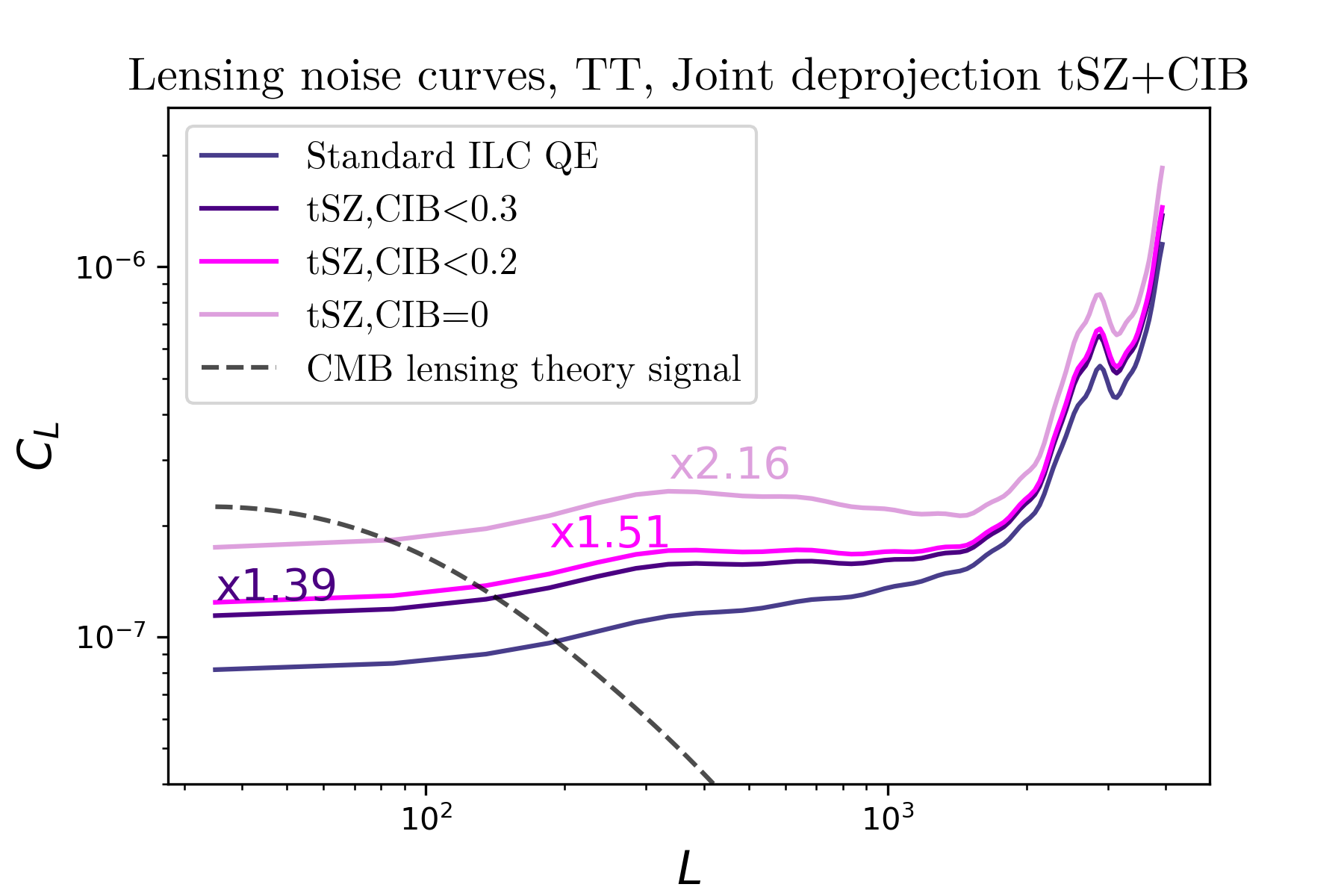}
\label{fig:sfig1joint}}
\qquad
\subfloat[Subfigure 2 caption:][Case with $\ell_{\rm{max}}=3500$ for reconstruction with the multi-frequency estimators.]{
 \includegraphics[width=0.46\linewidth]{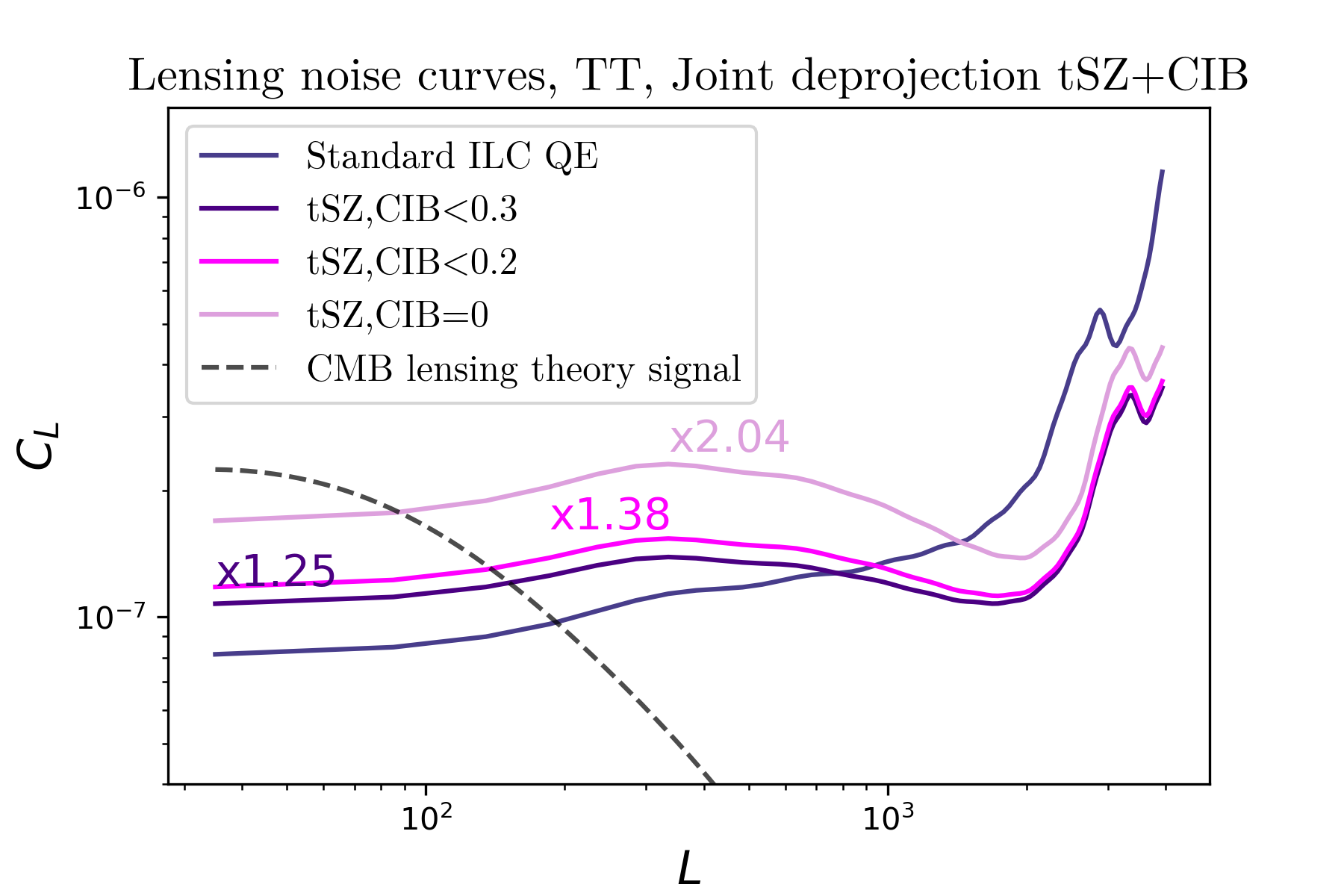}
\label{fig:sfig2joint}}
\caption{(Partial) deprojection of both tSZ and CIB in CMB maps used for lensing reconstruction, using the temperature CMB noise curves in Figure \ref{FIG:tw}. In both panels (a) and (b) the lensing noise curves are shown for four cases: Standard ILC CMB maps used in a quadratic lensing estimator; a CMB map cleaned with cILC, used in a quadratic lensing estimator; and pcILC-cleaned CMB maps, again used in a quadratic lensing estimator. The multi-frequency-cleaned maps are labelled with a number that represents the average increase in noise with respect to lensing reconstruction using the standard ILC CMB map. It can be seen that the pcILC gives a significant noise reduction on large scales of around $30\%$, when compared with cILC foreground mitigation methods.}
\label{fig:figjoint}
\end{figure*}

\begin{figure*}
\centering
\subfloat[Subfigure 1 caption:][Case with $\ell_{\rm{max}}=3000$ for reconstruction with all estimators.]{
\includegraphics[width=0.46\linewidth]{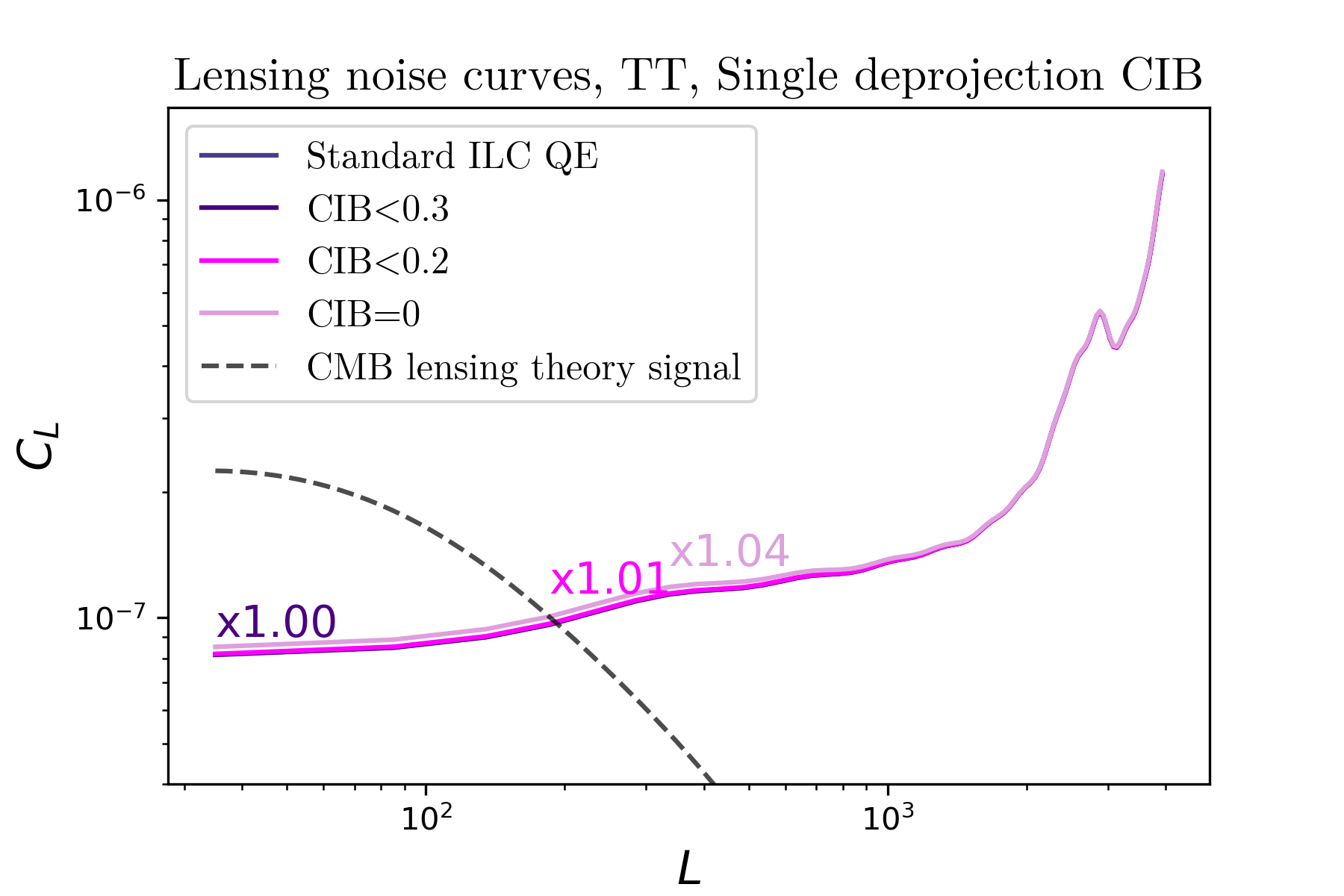}
\label{fig:sfig1CIB}}
\qquad
\subfloat[Subfigure 2 caption:][Case with $\ell_{\rm{max}}=3500$ for reconstruction with the multi-frequency estimators.]{
 \includegraphics[width=0.46\linewidth]{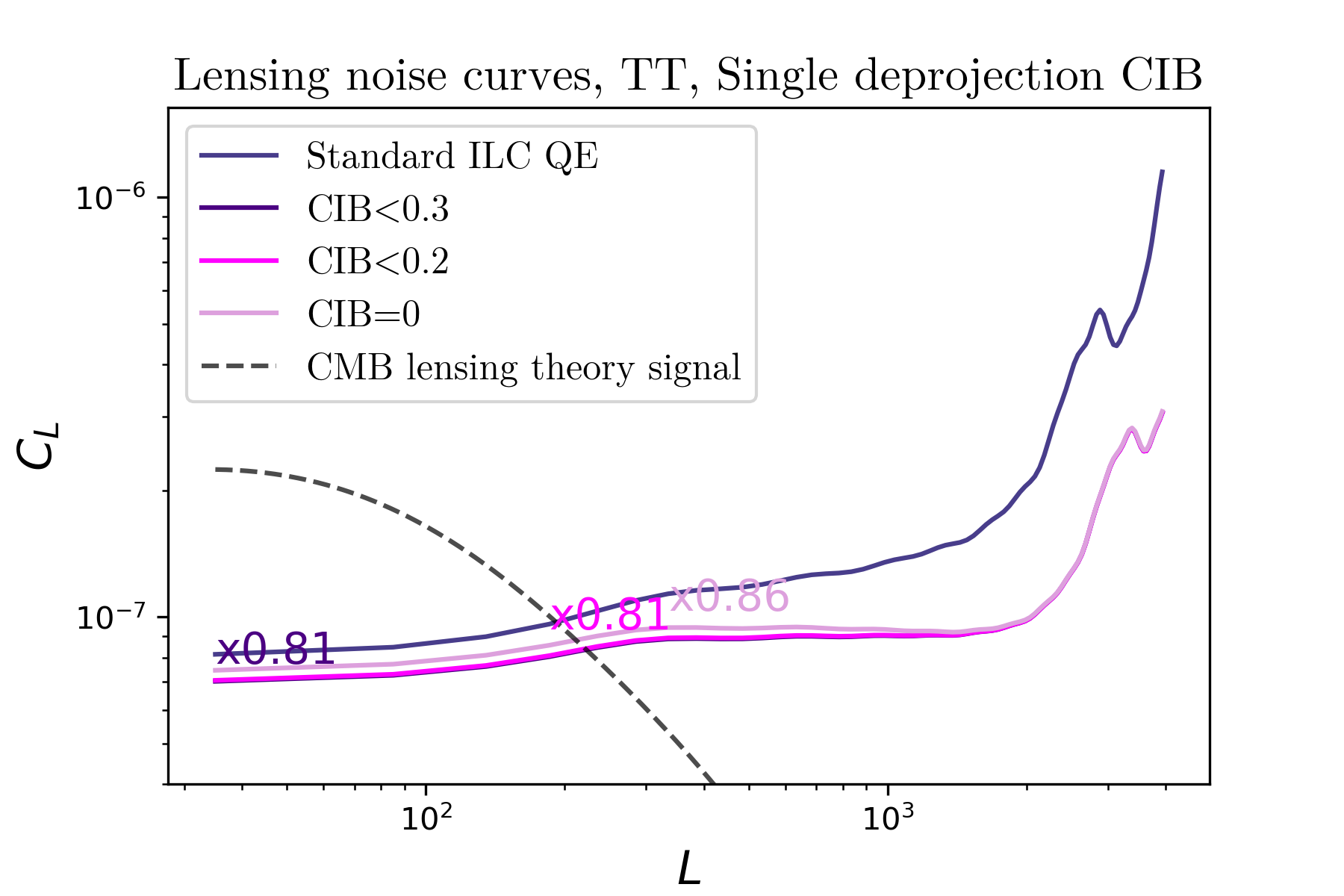}
\label{fig:sfig2CIB}}
\caption{As for Figure~\ref{fig:figjoint}, but (partially) deprojecting only CIB, using the temperature CMB noise curves in Figure \ref{FIG:ci}. In this case there are not relevant CMB lensing noise improvements between cILC and pcILC, as CIB deprojection does not lead to a huge blowing up in CMB temperature noise for the CMB scales of lensing reconstruction.}
\label{fig:figCIB}
\end{figure*}

\begin{figure*}
\centering
\subfloat[Subfigure 1 caption:][Case with $\ell_{\rm{max}}=3000$ for reconstruction for all estimators.]{
\includegraphics[width=0.46\linewidth]{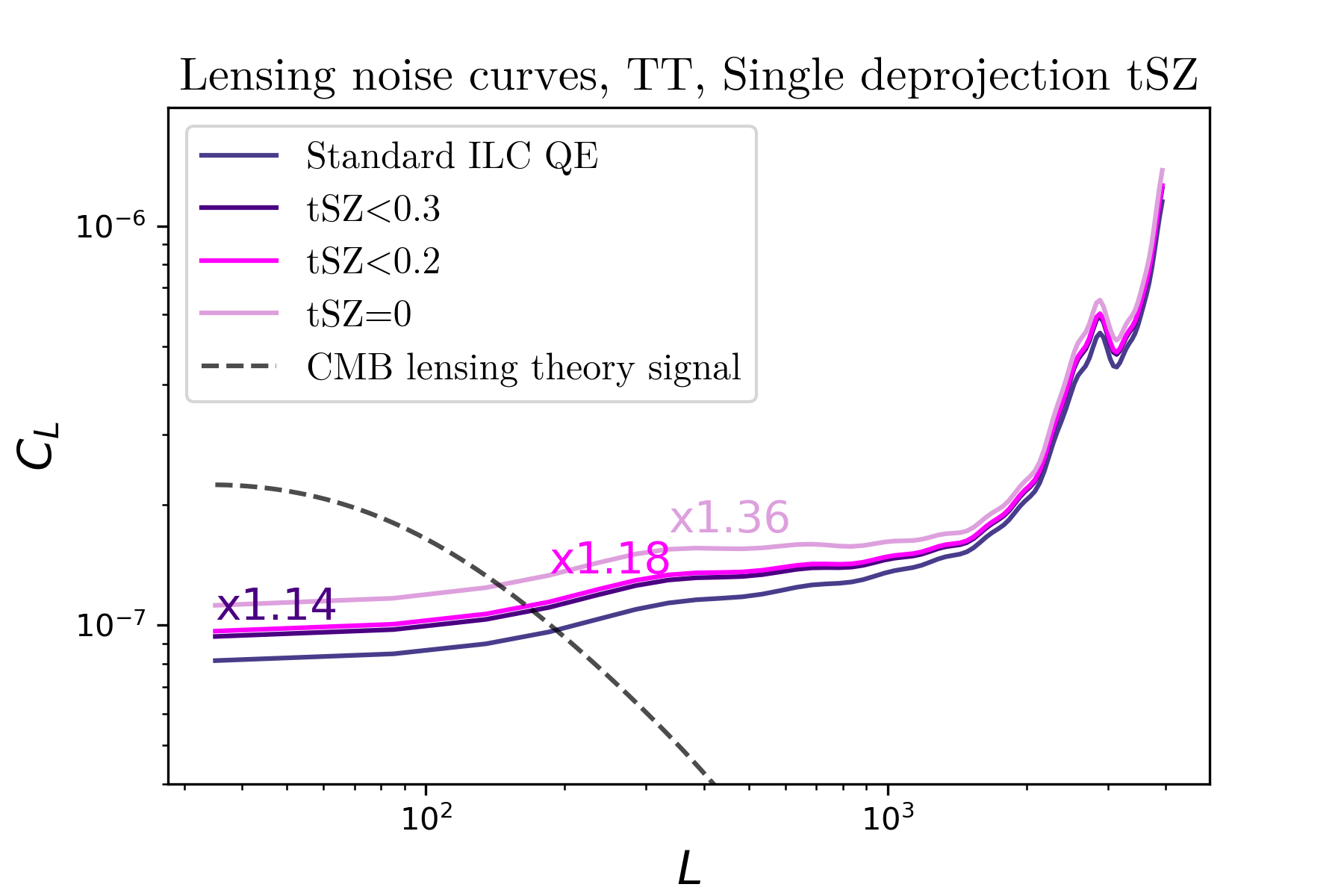}
\label{fig:sfig1tSZ}}
\qquad
\subfloat[Subfigure 2 caption:][Case with $\ell_{\rm{max}}=3500$ for reconstruction with the multi-frequency estimators.]{
\includegraphics[width=0.46\linewidth]{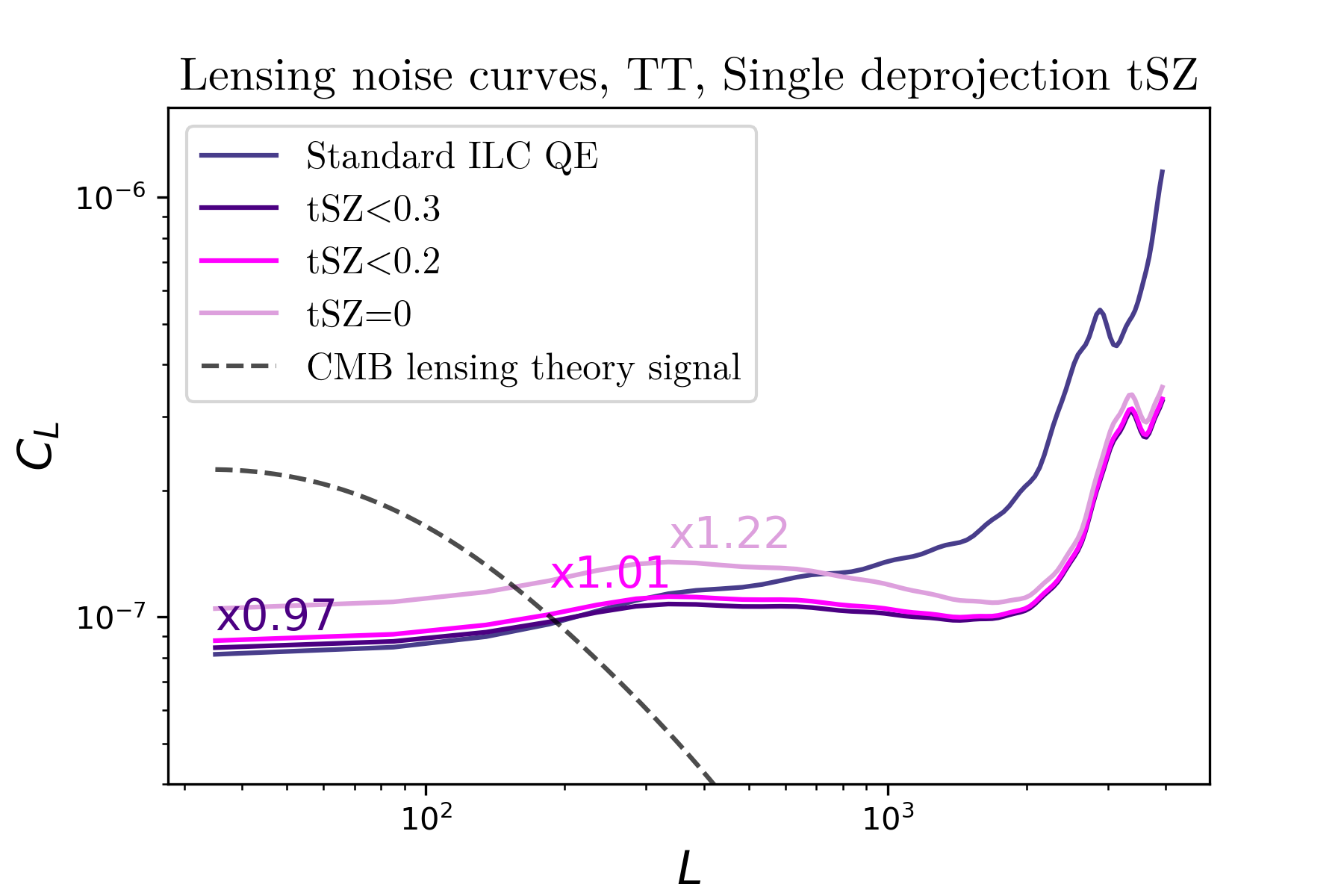}
\label{fig:sfig2tSZ}}
\caption{As for Figure~\ref{fig:figjoint}, but (partially) deprojecting only tSZ, using the temperature CMB noise curves in Figure \ref{FIG:sz}. On large scales, the pcILC derived CMB lensing noise performs better than cILC one at around $10\%$. }
\label{fig:figtSZ}
\end{figure*}

\subsection{Application to CMB lensing reconstruction}
In this subsection, we will present, as an example, an application of our new pcILC method for foreground mitigation: foreground reduction for CMB lensing analysis.

Along their paths to our telescopes, CMB photons are deflected, or lensed, by the gravitational influence of matter in our Universe. CMB lensing measurements allow us to constrain key cosmological parameters, such as the equation of state of dark energy, the sum of neutrino masses, or the amplitude of density fluctuations (e.g.,~\cite{2017PhRvD..95l3529S,2019ApJ...884...70W,2020A&A...641A...8P}).

Thanks to high-resolution, low-noise CMB surveys, it is possible to extract the CMB lensing signal with quadratic estimators, exploiting the lensing-induced couplings between different modes of the CMB (e.g., \cite{Hu_2002}). However, these estimators are susceptible to the presence of foreground contamination in mm-wave maps, leading to potential biases in the extracted cosmological parameters~\cite{vanEngelen2014,Osborne_2014,Ferraro_Hill_2018,Schaan_2019}.  Foregrounds are a more significant limitation for CMB temperature-derived lensing reconstruction than for polarization-derived reconstruction, as small-scale foregrounds in polarization are smaller compared to the CMB signal~\cite{Beck_2020}.  If CMB temperature foregrounds are left untreated, the resulting CMB lensing auto-spectrum and cross-spectrum analyses may be biased at the 3-20\% levels, much larger than the statistical error bars (e.g.,~\cite{vanEngelen2014,Ferraro_Hill_2018,Madhavacheril_2018,Omori2019,Baxter2019,Sailer2020}).  As many current- and next-generation lensing maps will still depend to a large extent on temperature data, rather than on polarization, it is important to mitigate these foregrounds for lensing analyses.

To mitigate foreground contamination in CMB lensing maps, different methods exist; these broadly divide into geometrical (e.g.,~\cite{Namikawa_2013,Osborne_2014,Schaan_2019}) and multi-frequency methods (e.g., \cite{Madhavacheril_2018,darwish2020atacama,Beck_2020}).  Here we will focus on the latter, although it is worth keeping in mind that this approach cannot mitigate foreground biases with the same SED as the primary CMB~\cite{Ferraro_Hill_2018}.  In particular, we will use the multi-frequency symmetric cleaned estimator presented in Ref.~\cite{darwish2020atacama} (building on the estimator of Ref.~\cite{Madhavacheril_2018}), which has the advantage of degrading the S/N of a CMB lensing measurement less than many other methods, while still significantly reducing the foreground-induced contamination.

In Ref.~\cite{darwish2020atacama} this method was applied to temperature tSZ-deprojected cILC maps, constructed using data from the Atacama Cosmology Telescope and the Planck satellite~\cite{2020PhRvD.102b3534M}. This lead to a reduction of the tSZ contamination in the reconstructed CMB lensing map, albeit at the price of somewhat decreased S/N in the measurement, when compared to the case without any deprojection. For current CMB lensing goals, a complete nulling of the foreground contamination is likely unnecessary; given that foreground biases are already quite small (a few percent of the signal), only a substantial reduction of the CMB lensing foreground contamination is likely required in order for it to be negligible. The pcILC method allows for the required reduction in the foreground contamination while improving the S/N relative to complete deprojection with a cILC.

We illustrate this S/N improvement from the use of the pcILC in Figures~\ref{fig:figjoint},~\ref{fig:figCIB}, and~\ref{fig:figtSZ}, where temperature CMB lensing reconstruction noise curves arising from the ILC, cILC, and pcILC temperature CMB maps are shown. We apply the standard quadratic estimator to the ILC map, and apply the method of Ref.~\cite{darwish2020atacama} to the cILC and pcILC maps. 
The CMB modes we use for reconstruction have $\ell_{\rm{min}}=30$; in addition, we use $\ell_{\rm{max}} = 3000$ for all the CMB lensing estimators; we also show results with $\ell_{\rm{max}} = 3500$ for the multi-frequency cleaned ones. Figure \ref{fig:figjoint} shows temperature estimator CMB lensing noise curves when constraining both tSZ and CIB (from Figure \ref{FIG:tw}), Figure \ref{fig:figCIB} shows constraints on only CIB (from Figure \ref{FIG:ci}), and Figure \ref{fig:figtSZ} only on tSZ (from Figure \ref{FIG:sz}).
It is clear that in terms of CMB lensing noise, the pcILC outperforms the cILC by a 
factor of around $10-30 \%$, even though the foreground bias is expected to be reduced to a level that is negligible for current and upcoming experiments. Multi-frequency cleaning with the pcILC method is therefore a powerful technique for mitigation of foreground biases in CMB lensing, while minimizing degradation in signal-to-noise.

\section{\label{sec:concl}Conclusions}

In this work, we have developed a new tool for CMB foreground cleaning -- the partially constrained ILC method (pcILC). This method finds the minimum-variance linear combination of different frequency channels in cases where residual foreground biases must be controlled to be below a threshold value, but do not need to strictly be nulled. By allowing for, in many cases, negligibly small but non-zero foreground residuals, this method provides significant reductions in variance -- often by factors of 2-3 -- when compared with a constrained ILC in which foregrounds are strictly nulled. We test and validate our method using realistic SO-like simulations, finding that we can reproduce the expected, forecast performance. Our method can be easily applied to current and upcoming CMB surveys, and has several possible applications; as an example, we show that it is capable of mitigating foreground biases in CMB lensing at lower noise than previous multi-frequency methods.

\begin{acknowledgments}
We thank Mathieu Remazeilles for useful discussions. SA acknowledges support from the Energetic Cosmos Laboratory.  BDS acknowledges support from a European Research Council (ERC) Starting Grant under the European Union’s Horizon 2020 research and innovation programme (Grant agreement No. 851274) and from an STFC Ernest Rutherford Fellowship. JCH thanks the Simons Foundation for support.  OD thanks the STFC.
\end{acknowledgments}

\section*{Appendix A: CIB bias}

In this section, we discuss why we find problems with the CIB bias obtained from the pcILC and cILC methods implemented for the simultaneous deprojection of two components. 

\begin{figure*}
\includegraphics[width=0.98\textwidth]{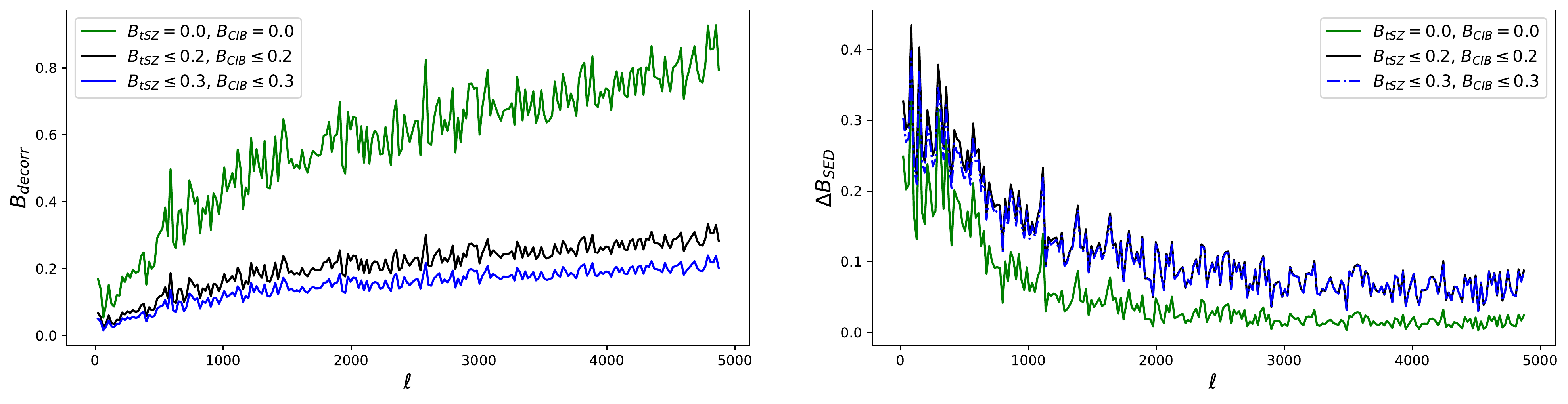}
\caption{Effect of decorrelation between frequency maps (right) and suboptimal CIB spectral response model (left) on CIB bias.}
\label{FIG:decorr}
\end{figure*}

As mentioned in Section~\ref{sec:results}, these problems arise due to slight decorrelation of the CIB across frequencies and the imperfect SED model of the CIB signal. To understand why this is so, let us first express the CIB cross-frequency power spectra as follows:
\begin{equation}
    F^{\text{cib}}_{\nu \times \nu'}=r^{\text{cib}}_{\nu \times \nu'}\sqrt{F^{\text{cib}}_{\nu \times \nu} F^{\text{cib}}_{\nu' \times \nu'}}
\end{equation}
where $r^{\text{cib}}_{\nu \times \nu'}$ is the correlation coefficient between $\nu$ and $\nu'$ frequency channels, and all auto- and cross-frequency spectra obtained are from the simulations.  Next, we split this equation into two parts, 
\begin{equation}
    F^{\text{cib}}_{\nu \times \nu'}=\sqrt{F^{\text{cib}}_{\nu \times \nu} F^{\text{cib}}_{\nu' \times \nu'}} + (r^{\text{cib}}_{\nu \times \nu'}-1)\sqrt{F^{\text{cib}}_{\nu \times \nu} F^{\text{cib}}_{\nu' \times \nu'}}
\end{equation}
where the first part is decorrelation-free and the second is the decorrelation estimate. Since the off-diagonal values of the covariance matrix $\mathbf{F}^{\text{cib}}$ consist of cross-frequency spectra, by analogy we can express the covariance matrix as follows:
\begin{equation}
    \mathbf{F}^{\text{cib}}=\mathbf{q}\mathbf{q}^T+\mathbf{q}\mathbf{q}^T \circ (\mathbf{r}-\mathbf{1})=\mathbf{Q}+\mathbf{D}
\end{equation}
where $\mathbf{q}^T=\bigg[\sqrt{F^{\text{cib}}_{\nu_1 \times \nu_1}}$ $\sqrt{F^{\text{cib}}_{\nu_2 \times \nu_2}} \cdots \sqrt{F^{\text{cib}}_{\nu_{N_{\nu}} \times \nu_{N_{\nu}}}}\bigg]$, $\mathbf{r}$ is the $N_{\nu}\times N_{\nu}$ matrix where each element corresponds to the correlation coefficient between frequency channels, and $\mathbf{1}$ is the $N_{\nu}\times N_{\nu}$ matrix where each element is equal to one. Then, using Eq. (\ref{bias_eq}), the CIB bias can be decomposed as follows: 
\begin{equation}
    B_{CIB}=\frac{\mathbf{w}^T\mathbf{Q}\mathbf{w}+\mathbf{w}^T\mathbf{D}\mathbf{w}}{F^{\text{cib}}_{145\times 145}}  
    \label{bias_split}
\end{equation}
The first part of Eq. (\ref{bias_split}) corresponds to the case where the CIB maps are perfectly correlated, so we can use it as a SED test:
\begin{equation}
    \Delta B_{SED}=\frac{\mathbf{w}^T\mathbf{Q}\mathbf{w}}{F^{\text{cib}}_{145\times 145}} - \frac{\mathbf{w}^T \mathbf{b} \mathbf{b}^T \mathbf{w}}{b^2_{145}} 
\end{equation}
where $\mathbf{b}$ is the model CIB SED (see Eq.~(\ref{eq: SED_CIB})), an $N_{\nu}\times 1$ vector, and $b_{145}$ is the model CIB SED evaluated at $145$ GHz. In Figure~\ref{FIG:decorr} (left), we show how the suboptimal CIB SED affects the CIB bias error.

\begin{figure*}
\includegraphics[width=0.98\textwidth]{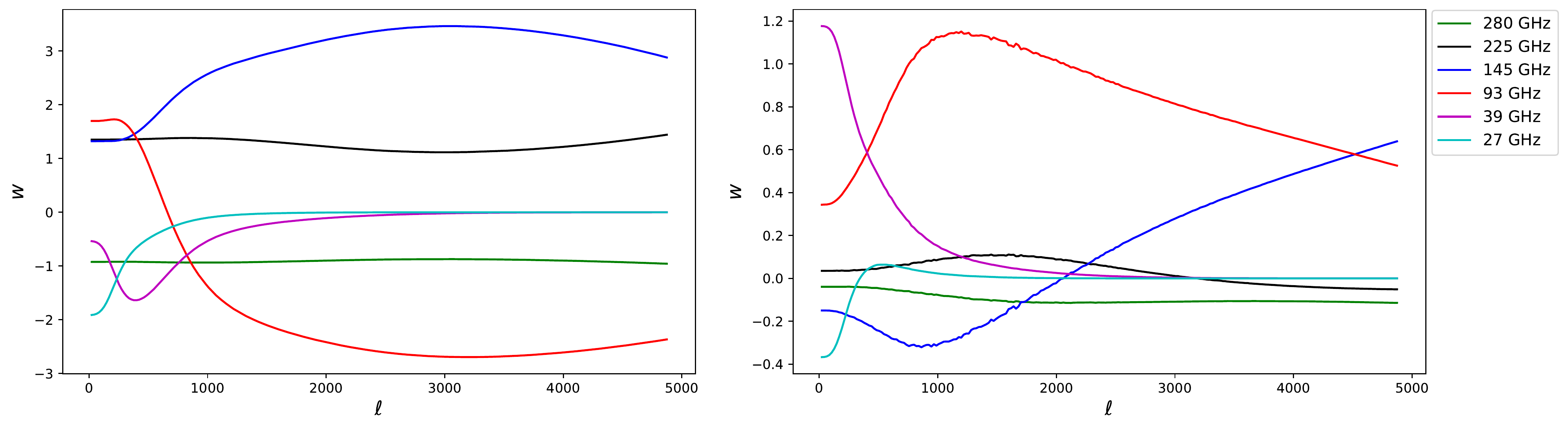}
\caption{Weight values for each frequency channel. Left: weights obtained from cILC for two-component (CIB and tSZ) deprojection. Right: weights obtained for CIB deprojection.}
\label{FIG:weights}
\end{figure*}

The second part of the Eq. (\ref{bias_split}) shows how decorrelation affects the CIB bias:
\begin{equation}
    B_{decorr}=\frac{\mathbf{w}^T\mathbf{D}\mathbf{w}}{F^{\text{cib}}_{145\times 145}}  
\end{equation}
In Figure~\ref{FIG:decorr} (right) we can see how decorrelation affects the CIB bias error.  As expected, the decorrelation errors grow with $\ell$, as the CIB decorrelation itself does~(e.g.,~\cite{Mak_2017}).

In Figure \ref{FIG:ci} (lr), we can see that the CIB bias residual when reducing or removing only one component is much lower compared with the CIB bias when reducing or removing two components. Since Eq.~(\ref{bias_split}) also applies for single-inequality ILC, the only difference is in the weights, which are shown in Figure~\ref{FIG:weights}.  An explanation for the imperfect CIB removal is therefore that the high weight values when deprojecting two components amplify small errors in the CIB model and small amounts of decorrelation, leading to significant biases.

\section*{Appendix B: Multiple Partially Constrained ILC}

Suppose we have $N_{\mathrm{f}}$ frequencies, $N_{\mathrm{c}}$ components, with $N_{\mathrm{c}}-1$ foregrounds. 
Suppose we would like to constrain $P$ foregrounds.

Let us define a few quantities: consider a ``projection'' operator $M$ described by a $2P\times P$ matrix,

\begin{equation}
    M \equiv \begin{pmatrix}
1 & 0 & 0 & 0 & ... &0 & 0\\
1 & 0 & 0 & 0 & ...&0 & 0\\
0 & 1 & 0 & 0 & ...&0 & 0\\
0 & 1 & 0 & 0 & ...&0 & 0\\
...& ...&...&...&... &... & ...\\
0&0&0&0&...&...&1\\
0&0&0&0&...&...&1
\end{pmatrix}
\end{equation}

and another $2P\times P$ operator $N$,
\begin{equation}
    N \equiv \begin{pmatrix}
1 & 0 & 0 & 0 & ... &0 & 0\\
-1 & 0 & 0 & 0 & ...&0 & 0\\
0 & 1 & 0 & 0 & ...&0 & 0\\
0 & -1 & 0 & 0 & ...&0 & 0\\
...& ...&...&...&... &... & ...\\
0&0&0&0&...&...&1\\
0&0&0&0&...&...&-1
\end{pmatrix}
\end{equation}

Also define a $2P$-dimensional vector containing the ``slack'' variables,
\begin{equation}
    \vec{s} \equiv \begin{pmatrix}
    s_{11}&
    s_{12}&
    ...&
    ...&
s_{j1}&
s_{j2}&
    ...&
    ...&
    s_{P1}&
    s_{P2}
    \end{pmatrix}^T
\end{equation}
and another one for their squares
\begin{equation}
    \vec{s}_2 \equiv \begin{pmatrix}
    s_{11}^2&
    s_{12}^2&
    ...&
    ...&
s_{j1}^2&
s_{j2}^2&
    ...&
    ...&
    s_{P1}^2&
    s_{P2}^2
    \end{pmatrix}^T
\end{equation}
where $j$ refers to the $j$-th foreground to be deprojected.

Note that we can write this vector as
\begin{equation}
    \vec{s}_2^T = \sum_{k} (\mathbf{P}_k\vec{s})^T(\mathbf{P}_k \vec{s})\vec{e}_k^T
\end{equation}
where $\vec{e}_k$ is an orthonormal basis vector and $\mathbf{P}_k$ is a projection matrix.

Also consider an $N_f\times P$ matrix defining the foregrounds to be constrained,
\begin{equation}
    F \equiv \begin{pmatrix}
    \vec{f}^1 & \vec{f}^2 &...&\vec{f}^P
    \end{pmatrix}
\end{equation}
with $\vec{f}^j$ the $N_{\mathrm{f}}$-dimensional vector where $(\vec{f}^j)_i$ is the foreground component $j$ SED at frequency $i$.

And finally define the constraints vector
\begin{equation}
\vec{\epsilon} \equiv    \begin{pmatrix}
    \epsilon_1 & \epsilon_2 & ...& \epsilon_P
    \end{pmatrix}^T
\end{equation}

Then we can write a Lagrangian
\begin{equation}
    \mathcal{L} \equiv \vec{w}^T R \vec{w} + \lambda (1-\vec{w}^T \vec{e})+\vec{\lambda}_{\mathrm{ineq}}^T M \vec{\epsilon} + \vec{\lambda}_{\mathrm{ineq}}^T N F^{T}\vec{w}-\vec{\lambda}_{\mathrm{ineq}}^T \vec{s}_2
\end{equation}

To solve for the weights, we then have to look at the surfaces of minimum functional
\begin{align}
\begin{cases}
  \vec{\nabla}_{\vec{w}^T}\mathcal{L} = 2 R \vec{w} - \lambda \vec{e} + FN^T\vec{\lambda}_{\mathrm{ineq}} = \vec{0}\\
   \frac{\partial\mathcal{L}}{\partial \lambda} = 1-\vec{e}^T\vec{w}=0\\
   \vec{\nabla}_{\vec{\lambda}_{\mathrm{ineq}}^T}\mathcal{L}=M\vec{\epsilon}+NF^T\vec{w}-\vec{s}_2=\vec{0}\\
   \vec{\nabla}_{\vec{s}^T}\mathcal{L}= -2\sum_k (\mathbf{P}^T_k \mathbf{P}_k \vec{s}) \vec{e}_k^T \vec{\lambda}_{\mathrm{ineq}} = \vec{0}
\end{cases}
\label{eq:generealcasesystem}
\end{align}

If the constraints are active, for the $\lambda_{\mathrm{ineq},jl}$ multiplier, then we have equality constraint, otherwise if the constraints are non-active, then we reduce to the standard ILC, as this is allowed.

\nocite{*}

\bibliographystyle{JHEP}
\bibliography{apssamp}

\end{document}